\documentclass[10pt,a4paper,twoside]{article}

\usepackage{amssymb,amsmath,multirow}

\newtheorem{theorem}{Theorem}[section]
\newtheorem{charact}[theorem]{Characterization}

\newtheorem{lemma}[theorem]{Lemma}

\usepackage[dvips]{graphicx}
\usepackage{epsfig}

\newcommand{\pa}{\mathcal{P}}
\newcommand{\lo}{\mathcal{L}}
\newcommand{\ex}{\mathcal{E}}

\newcommand{\integ}{\int\limits_1^\infty}

    \makeatletter
    \let\@fnsymbol\@arabic
    \makeatother

\title{Two-dimensional Kolmogorov-type Goodness-of-fit Tests Based on Characterizations and their Asymptotic Efficiencies}

\author{Bojana Milo\v sevi\'c, Marko Obradovi\'c}

\date{}
\begin{document}

\maketitle
\begin{abstract}
In this paper new  two-dimensional goodness of fit tests are proposed. They are of supremum-type and are based on different types of characterizations. 
For the first time a characterization based on independence of two statistics
 is used for goodness-of-fit testing. The asymptotics of the statistics is studied and Bahadur efficiencies of the tests against some close 
alternatives are calculated. In the process a theorem on large deviations of Kolmogorov-type statistics has been extended to the multidimensional case.
\end{abstract}



\section{Introduction}
Goodness of fit testing has for a long time been an important topic in statistics.
In recent times the approach of constructing tests based on characterizations of distributions has become very popular.
Starting from \cite{vasicek}, \cite{koul1} and \cite{koul2} there have been plenty of tests proposed  based on various types of characterizations. 
Some of them can be found in
\cite{angus}, \cite{baringhaus}, \cite{henzeMeintanis}, \cite{ahmad}, \cite{morris}, etc.

A large portion of these tests are based on empirical distribution functions and their comparison in some norm.
A notable representative is the test using the supremum ($L^{\infty}$) norm, so-called Kolmogorov-type test.

For purpose of measuring the quality of test an important tool is its asymptotic efficiency.
Since the limiting distribution of the Kolmogorov-type statistics is not Gaussian, the classical Pitman approach to measure the asymptotic
efficiency is not applicable. Therefore the Bahadur efficiency has emerged as a natural choice. Its popularity has been increased with the development of
large deviation theory. For a large deviation results of Kolmogorov-type statistics see \cite{nikitinLDKS}.

The Bahadur efficiency of some tests based on characterizations has been studied in many papers (see e.g. \cite{nikitinDurio}, \cite{nikitinGeorgian}, \cite{jovanovic},
\cite{obradovic}, \cite{NikitinHenze}, \cite{litvinova}, \cite{volkovaRossberg}, \cite{volkovaShepp}, \cite{volkovaPower}).
In all of them except \cite{tchirina} the supremum is considered over a subset of the real line, i.e. over one dimensional set.

Here we propose some tests that have supremum statistic over two dimensional set (subset of $R^2$). When constructing goodness of fit tests based on characterizations,
this situation naturally arises in the following circumstances.
For example, consider a characterization of univariate distribution based on a functional equation with two parameters. A famous example is the lack of memory property
\begin{equation*}
1-F(x+y)=(1-F(x))(1-F(y)),
\end{equation*}
that characterizes the exponential distribution. A test based on this property was examined by Tchirina (\cite{tchirina}).

Another possibility is the
characterization of univariate distribution based on independence of two statistics. Besides, such test statistics
may appear when characterizing multivariate distributions or even when constructing standard goodness of fit tests for multivariate distributions.
If the components of this multivariate distribution are independent such testing is also closely related to testing the hypotheses of independence.

In this paper we study goodness of fit tests based on characterization of two types, namely functional equations and the independence of two statistics. 
The functional equations type tests have been considered before, while this is the first goodness of fit test based on independence characterization.

The paper is organized as follows. In section 2 we present large deviation theorem for supremum type statistics. 
In section 3 we present the characterizations and test statistics. Their local Bahadur efficiencies are presented in section 4.

\section{Supremum-type Statistics and their Large Deviations}
In this section we present a theorem which extends the result od Nikitin (\cite{nikitinLDKS}) to cover the multidimensional parameter case.

Since our test statistics are based on some $U$-statistics we present the following theorem on large deviation which is stated and proved in \cite{nikitinPonikarov}.
\begin{theorem}
Consider the U-statistic of degree $m \geq 1$
\begin{equation}
U_n=\frac{1}{\binom{n}{m}}\sum\limits_{1\leq i_1<\cdots< i_m\leq n}\Phi(X_{i_1},...,X_{i_m})
\end{equation}
with centered, bounded, and non-degenerate real-valued kernel $\Phi$ so that
\begin{equation*}
E\Phi(X_1,...,X_n)=0,\;\;|\Phi(x_1,...,x_n)|\leq M,
\end{equation*}
and $\sigma^2=E\varphi^2(X_1)>0$ with $\varphi(y)=E(\Phi(X_1,...,X_n)|X_1=y)$. Than we have
\begin{equation}\label{gfi}
\lim\limits_{n \to \infty}\frac{1}{n}\log P(U_n\geq \varepsilon):=-g_{\Phi}(\varepsilon)=-\sum\limits_{j=2}^{\infty}b_j\varepsilon^j,
\end{equation}
where the series converges for sufficiently small $\varepsilon>0$ and $b_2=\frac{1}{2m^2\sigma^2}$.
\end{theorem}

For more theory on U-statistics we refer to \cite{korolyuk} and \cite{serfling}.

\bigskip

Consider the statistics of the form
\begin{equation*}
K_n=\sup\limits_{\mathbf{t}\in T}\big|U_n(\mathbf{X}_m,\mathbf{t})\big|,\end{equation*}
where $\{U_n(\mathbf{t}), \; \mathbf{t} \in T\}$ is a family of U-statistics of order $m$ with centered, bounded, non-degenerate kernel $\Phi$, and $T=[a_1,b_1]\times\cdots\times[a_p,b_p]\subset R^p$.
Besides, we suppose that the family $U_n(\mathbf{t})$ is non-degenerate, i.e. its variance function $\sigma^2_{\varphi}(\mathbf{t})=E(\varphi^2(X_1,\mathbf{t}))>0$ for $t$ in interior of $T$ except in finite number of points. Denote
\begin{equation*}
\sigma^2_0=\sup\limits_{\mathbf{t}\in T}\sigma_{\varphi}^2(t).
\end{equation*}

 Moreover, we suppose that the family $U_n(\mathbf{t})$ satisfies condition  called \textit{monotonicity in parameter}. We generalize its definition from \cite{nikitinLDKS} to multidimensional case.
 Suppose there exist sequences of partitions $\mathcal{A}_N$ of the set $T$
 \begin{equation}
 \mathcal{A}_N=\Big\{\prod\limits_{i=1}^p[t^{(i)}_{k_i},t^{(i)}_{k_{i+1}}], \;k_i=0,..N-1,\; i=1,...,p\Big\}
 \end{equation}
such that the nodes of
 the partition do not coincide with zeros of variance function and for any $i$ and $k_i$  holds
 \begin{equation}
 \sup\limits_{\mathbf{t}\in \prod\limits_{i=1}^p[t^{(i)}_{k_i},t^{(i)}_{k_{i+1}}]}U_n(\mathbf{ t})\leq U_n(t^{(1)}_{k_1+1},...,t^{(p)}_{k_p+1})+\Delta_n(N),
 \end{equation}
where the sequence $\Delta_n(N)$ converges  to zero rapidly enough. That is, there exist the sequence $\tau_N$ which converges to zero when $N\to \infty$ such that
\begin{equation*}
\lim\limits_{N\to \infty}\lim\limits_{n \to \infty}\frac{1}{n}\log P\{\Delta_n(N)>\tau_N\}=-\infty.
\end{equation*}

\begin{theorem}\label{glavna}
Let $\{U_n(\mathbf{t}), t \in T\}$ be non-degenerate family of $U$-statistics with kernels $\Phi(\mathbf{x},\mathbf{t})$ which is bounded and centered for all $t$ and satisfies the condition monotonicity in parameter. Than for  $\varepsilon >0$ the following limit holds true
\begin{equation}
\lim\limits_{n \to \infty}\frac{1}{n}\log P\{K_n\geq \varepsilon\}=-\inf\limits_{\mathbf{t}\in T}g_\phi(\varepsilon,\mathbf{t}),
\end{equation}
where $g_\phi(\varepsilon,\mathbf{t})$ is defined in \eqref{gfi}. For sufficiently small $\varepsilon$ this limit  can be represented as
\begin{equation}
g_K(\varepsilon):=\inf\limits_{\mathbf{t}\in T}g_\phi(\varepsilon,\mathbf{t})=\frac{\varepsilon^2}{2m^2\sigma^2_0}+O(\varepsilon^3),\; \varepsilon \to 0.
\end{equation}
\end{theorem}
This theorem is a generalization of (\cite{nikitinLDKS}, theorem 2.3).  The proof is analogous so we omit it here.

The test statistics we are going to study in this paper will be two-dimensional, and based on some characterizations of univariate distribution. For example, consider  a characterization based on independence of two functions of random variables.
Let $X_1,X_2,...,X_m$ be i.i.d. random variables from a distribution $F$ and let the independence of
$\omega_1(X_1,...,X_m)$ and $\omega_2(X_1,...,X_2)$ characterize the distribution $F$. Let $G_1$, $G_2$ and $H$ be their marginal and common distribution functions,
respectively.
The characterization then can be expressed as
\begin{equation*}
H(t_1,t_2)=G_1(t_1)G(t_2),\;\; t_1, t_2\in R.
\end{equation*}
Using U-empirical distribution functions we can make the  following test statistics
\begin{equation}\label{supremumopsta}
\sup\limits_{t_1,t_2}|H_n(t_1,t_2)-G_n(t_1,t_2)|,
\end{equation}
where $G_n(t_1,t_2)$ is product of marginal empirical  distribution functions. Here we take $H_n(t_1,t_2)=\sum I\{\omega_1(\textbf{X})<t_1,\omega_2(\textbf{X})<t_2\}$. 
It is known  that in two dimensions the distribution functions, and therefore the empirical d.f.'s are not uniquely defined. For example $H_n(t_1,t_2)$ can be defined as 
 $\sum I\{\omega_1(\textbf{X})>t_1,\omega_2(\textbf{X})<t_2\}$. This may cause some incoveniences when constructing multivariate tests (see e.g. \cite{fasano}).
However, our test statistic \eqref{supremumopsta} is invariant to any choice of distribution function.

\section{The Characterizations and the Tests}

We now present three characterizing theorems of three different distributions.
The first two characterizations, of the functional equation type, for Pareto and logistic distributions, can be found in \cite{galambos}.

\begin{charact}\label{chPar}
 Let the distribution function $F$ of the random variable $X$ be continuous non-negative random variable. Then the following equality
 \begin{equation*}
      1-F(xy)= (1-F(x))(1-F(y)) \text{ for } x,y>1,
      \end{equation*}
      holds if and only if $X$ follows Pareto distribution with the distribution function $F(x)=1-x^{-\lambda},\;\;x>1,\lambda>0$.
\end{charact}

  \begin{charact}\label{chLog}
   Let the distribution function $F$ of the random variable $X$ be continuous and symmetric about the origin. Then the following equality
     \begin{equation*}
      \frac{1-F(x+y)}{(1-F(x))(1-F(y))}= \frac{F(x+y)}{F(x)F(y)}
     \end{equation*}
     holds if and only if $X$ follows logistic distribution with the distribution function
     \begin{equation}\label{logistic}
      F(x)=\frac{1}{1+e^{-\lambda x}},\;x\in \mathbb{R},\;\lambda>0.
     \end{equation}
  \end{charact}

The third characterization, of independence type, can be found in \cite{fisz}.
\begin{charact}\label{chExp}
 If $X$ and $Y$ are i.i.d. random variables with an absolutely continuous distribution and if $\min\{X,Y\}$ and $|X-Y|$ are independent, then both $X$ and
$Y$ have exponential distribution with distribution function $F(x)=1-e^{-\lambda x},\;\;x>0,\;\lambda>0$.
\end{charact}

Next we shall present the tests based on the above characterizations and examine their asymptotics.

\subsection{Goodness-of-fit Test for Pareto Distribution}

Let $X_1,\ldots,X_n$ be a sample from non-negative continuous distribution $F$. In order to test the composite null hypothesis $H_0$ that the sample is from Pareto distribution
we use the following test statistic 

\begin{equation*}
 K_n^{\pa}=\sup\limits_{t_1,t_2>1}|H_n^{\pa}(t_1,t_2)-G_n^{\pa}(t_1,t_2)|,
\end{equation*}
where

\begin{equation*}
 H_n^{\pa}(t_1,t_2)=\frac{1}{\binom{n}{2}}\sum\limits_{1\leq i_1<i_2\leq n}\frac{1}{2!}\sum_{\pi(1:2)}I\{X_{i_{\pi(1)}}> t_1\}I\{X_{i_{\pi(2)}}> t_2\}
 \end{equation*}
 and
 \begin{equation*}
 G_n^{\pa}(t_1,t_2)=\frac{1}{n}\sum\limits_{i=1}^nI\{X_i>t_1t_2\},
 \end{equation*}
 are symmetrized $U$-empirical distribution functions and $\pi(1:r)$ represents the set of all $r!$ permutations of the set $1,\ldots,r$.
 
 If $X$ follows Pareto distribution with shape parameter $\lambda$ than $X^{\lambda}$ follows  Pareto distribution with shape parameter 1. Since the test statistic is 
 invariant to such transformations we can take without loss of generality $\lambda=1$. 
 
\begin{figure}[h!]
\begin{center}
\includegraphics[scale=0.8]{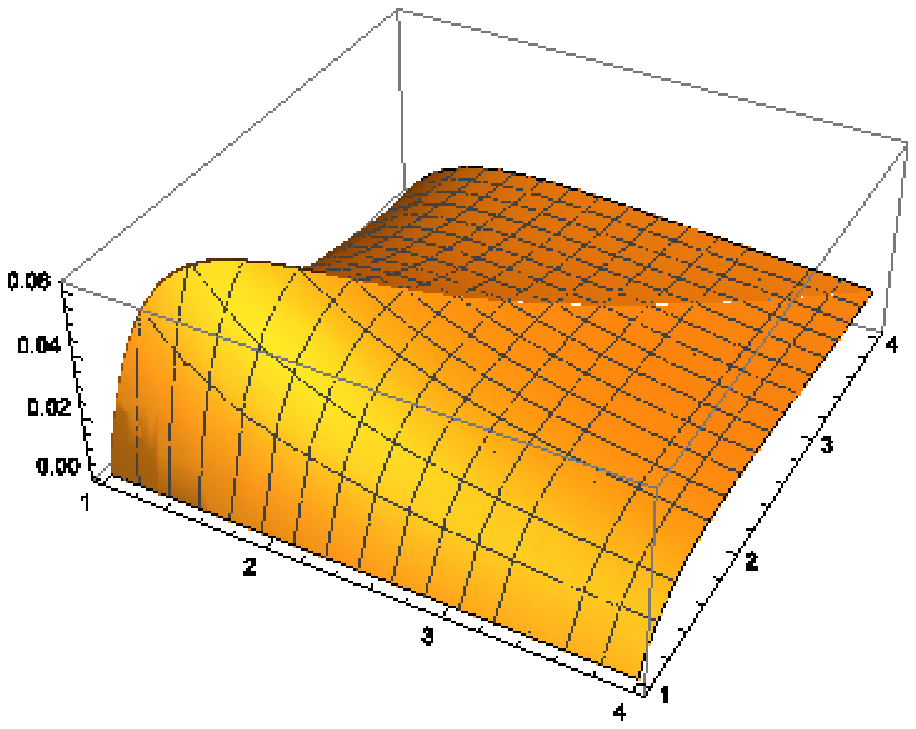}\caption{Plot of the function $\sigma_{\pa}^2(t_1,t_2),$  }
\label{fig: sigmaK2}
\end{center}
\end{figure}

 For a fixed $t_1>1$ and $t_2>1$ the expression $H_n^{\pa}(t_1,t_2)-G_n^{\pa}(t_1,t_2)$  is a $U$-statistic with symmetric kernel
 \begin{align*}
 \Xi^{\pa}(X,Y,t_1,t_2)=&\frac{1}{2}I\{X>t_1\}I\{Y>t_2\}+\frac{1}{2}I\{X>t_2\}I\{Y>t_1\}\\-&\frac{1}{2}I\{X>t_1t_2\}
 -\frac{1}{2}I\{Y>t_1t_2\}.
 \end{align*}
 Its projection on $X$ is
 \begin{align*}
 \xi^{\pa}(s)=&E(\Xi^{\pa}(X,Y,t_1,t_2)|X=s)=-\frac{1}{2t_1t_2}+\frac{1}{2t_2}I\{s>t_1\}\\+&\frac{1}{2t_1}I\{s>t_2\}-\frac{1}{2}I\{s>t_1t_2\}.
 \end{align*}
 The variance of this projection is
 \begin{equation}
 \sigma_{\pa}^2(t_1,t_2)=\frac{-1 + t_1 - t_2 + t_1 t_2}{4 t_1^2 t_2^2}I\{t_1<t_2\}+\frac{-1 -t_1 + t_2 + t_1 t_2}{4 t_1^2 t_2^2}I\{t_1\geq t_2\}.
 \end{equation}
 The plot of this function is shown in Figure \ref{fig: sigmaK2}.

We find that
\begin{equation*}
\sigma_{\pa}^2=\sup\limits_{t1>1,t2>1}\sigma_{\pa}^2(t_1,t_2)=\sigma_{\pa}^2(1.414,1.414)=0.0625.
\end{equation*}
Therefore the family of our test statistics is non-degenerate.

Limiting distribution of the statistic $K_n^{\pa}$ is unknown, but it can be shown that the
$U$-empirical process
\begin{equation*}\nu^{\pa}_n(t_1,t_2) =\sqrt{n} \left(H_n^{\pa}(t_1,t_2) - G_n^{\pa}(t_1,t_2)\right), \ t_1>1,\,t_2>1
\end{equation*}
weakly converges in $D(1,\infty)$ as $n \to \infty$ to certain centered Gaussian field $\nu(t_1,t_2)$ (see \cite{ruymgaart}). Then the sequence of statistics
$\sqrt{n} K_n^{\pa}$ converges in distribution to the random variable   $\sup_{t_1\geq0,t_2\geq 0} |\nu(t_1,t_2)|$ but it is impossible to find explicitly its distribution.

\subsection{Goodness-of-fit Test for Logistic Distribution}

Let $X_1,\ldots,X_n$ be a sample from a real-valued continuous distribution $F$. In order to test the composite null hypothesis $H_0$ that the 
sample is from the logistic distribution
\eqref{logistic}
we use the following test statistic

\begin{equation*}
 K_n^{\lo}=\sup\limits_{t_1,t_2\in \mathbb{R}}|H_n^{\lo}(t_1,t_2)-G_n^{\lo}(t_1,t_2)|,
\end{equation*}
where

\begin{equation*}
 H_n^{\lo}(t_1,t_2)=\frac{1}{\binom{n}{3}}\sum\limits_{i_1<i_2<i_3}\frac{1}{3!}\sum_{\pi(1:3)}I\{X_{i_{\pi(1)}}<t_1+t_2\}I\{X_{i_{\pi(2)}}>t_1\}I\{X_{i_{\pi(3)}}>t_2\}
 \end{equation*}
 \begin{equation*}
 G_n^{\lo}(t_1,t_2)=\frac{1}{\binom{n}{3}}\sum\limits_{ i_1<i_2<i_3}\frac{1}{3!}\sum_{\pi(1:3)}I\{X_{i_{\pi(1)}}>t_1+t_2\}I\{X_{i_{\pi(2)}}<t_1\}I\{X_{i_{\pi(3)}}<t_2\}.
 \end{equation*}
 
 Since the statistic is invariant to sample transformations of the type $\lambda X$, we can take $\lambda=1$.
 
\begin{figure}[h!]
\begin{center}
\includegraphics[scale=0.8]{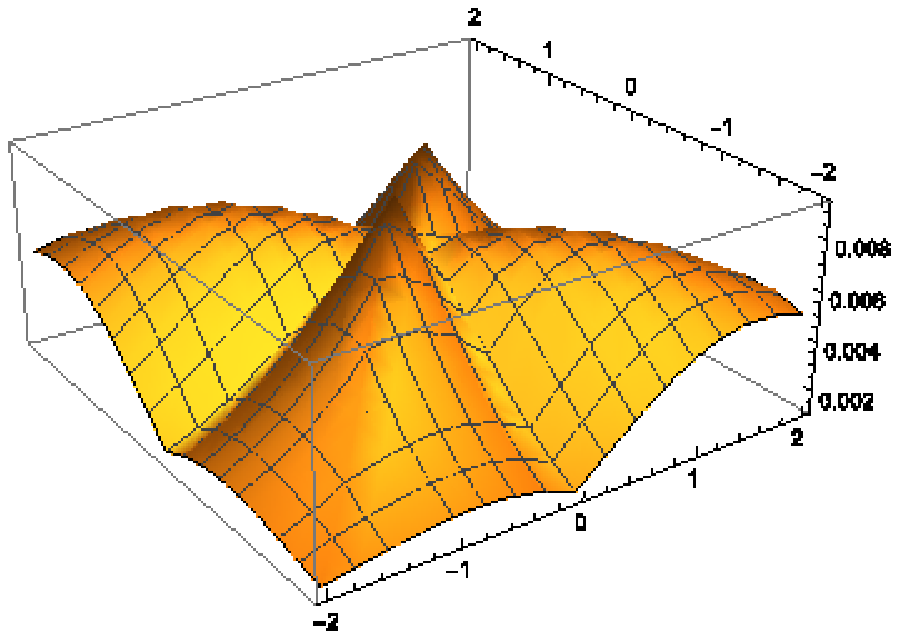}\caption{Plot of the function $\sigma_{\lo}^2(t_1,t_2),$  }
\label{fig: sigmaLog}
\end{center}
\end{figure}

 For a fixed $t_1>1$ and $t_2>1$ $H_n^{\lo}(t_1,t_2)-G_n^{\lo}(t_1,t_2)$  is a $U$-statistic with symmetric kernel
 \begin{align*}
 \Xi^{\lo}(X_1,X_2,X_3,t_1,t_2)=&\frac{1}{6}\sum\limits_{\pi(3)}\Big(I\{X_{i_1}<t_1+t_2\}I\{X_{i_2}>t_1\}I\{X_{i_3}>t_2\}
 \\-&I\{X_{i_1}>t_1+t_2\}I\{X_{i_2}<t_1\}I\{X_{i_3}<t_2\}\Big)
 \end{align*}
 Its projection on $X_1$ under $H_0$ is
 \begin{align*}
 \xi^{\lo}(s,t_1,t_2)=&E(\Xi^{\lo}(X_1,X_2,X_3,t_1,t_2)|X_1=s)=\frac{-1 + e^{t_1} + e^{t_2} + e^{t_1 + t_2}}{3 (1 + e^{t_1}) (1 + e^{t_2}) (1 + e^{t_1 + t_2})}
 \\-&\frac{e^{t_2} (1 + e^{t_1})}{3  (1 + e^{t_2}) (1 + e^{t_1 + t_2})}I\{s>t_1\}-\frac{e^{t_1} (1 + e^{t_2})}{3 (1 + e^{t_1})  (1 + e^{t_1 + t_2})}I\{s>t_2\}\\&+\frac{1+e^{t_1+t_2}}{3 (1 + e^{t_1}) (1 + e^{t_2}) }I\{s>t_1+t_2\}
 \end{align*}
 The expression for the variance function $\sigma_{\lo}^2$ is too complicated to display. Its plot is given in the Figure \ref{fig: sigmaLog}.
 The supremum of the variance function is
 \begin{equation*}
 \sigma_{\lo}^2=\sup\limits_{t_1,t_2}\sigma_{\lo}^2=\sigma_{\lo}^2(0.669,0.669)=0.00945.
 \end{equation*}
 Therefore the family of kernels $\Xi^{\lo}(\mathbf{X},t_1,t_2))$ is non-degenerate. Using similar arguments to the previous test statistic we can show that the
corresponding random process converges to a Gaussian process while the distribution of the  statistic $K_n^{\lo}$ remains unknown.
 \subsection{Goodness-of-fit Test for Exponential Distribution}

\begin{figure}[h!]
\begin{center}
\includegraphics[scale=0.8]{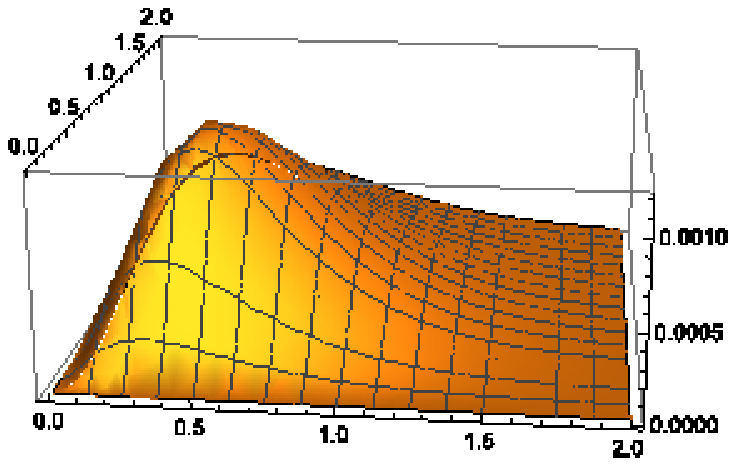}\caption{Plot of the function $\sigma_{\ex}^2(t_1,t_2),$  }
\label{fig: sigmaK1}
\end{center}
\end{figure}

Let $X_1,\ldots,X_n$ be a sample from non-negative continuous distribution $F$. In order to test the composite null hypothesis $H_0$ that the sample is from exponential distribution
we use the following test statistic free of scale parameter $\lambda$

\begin{equation*}
 K_n^{\ex}=\sup\limits_{t_1,t_2>0}|H_n^{\ex}(t_1,t_2)-G_n^{\ex}(t_1,t_2)|,
\end{equation*}
where
\begin{equation*}
 G^{\ex}_n(t_1,t_2)=\frac{1}{\binom{n}{2}}\sum\limits_{1\leq i<j\leq n}I\{\min(X_i,X_j)\leq t_1, |X_i-X_j|\leq t_2\},
\end{equation*}
\begin{equation*}
 H^{\ex}_n(t_1,t_2)=\frac{1}{\binom{n}{2}}\sum\limits_{1\leq i<j\leq n}I\{\min(X_i,X_j)\leq t_1\}\cdot\frac{1}{\binom{n}{2}}\sum\limits_{1\leq k<l\leq n}I\{|X_k-X_l|\leq t_2\}.
\end{equation*}
For a fixed $t_1>0$ and $t_2>0$ the expression $H^{\ex}_n(t_1,t_2)-G^{\ex}_n(t_1,t_2)$ is the $U$-statistic with the following kernel:
\begin{align*}
\Xi^{\ex}(X,Y,Z,W,t_1,t_2)&=\frac{1}{2}I\{\min(X,Y)\leq t_1\}\big(I\{|X-Y|\leq t_2\}-I\{|Z-W|\leq t_2\}\big)\\&+\frac{1}{2}
I\{\min(Y,Z)\leq t_1\}\big(I\{|Y-Z|\leq t_2\}-I\{|X-W|\leq t_2\}\big).
\end{align*}
The projection of this family of kernels on $X$ under $H_0$ is
\begin{align*}
\xi^{\ex}(s,t_1,t_2)&=E(\Xi^{\ex}(X,Y,Z,W,t_1,t_1)|X=s)=\frac{1}{2}P\{\min(Y,s)\leq t_1,|Y-s|\leq t_2\}
\\&-\frac{1}{2}P\{\min(Y,s)\leq t_1,|Z-W|\leq t_2\}
+\frac{1}{2}P\{\min(Y,Z)\leq t_1,|Y-Z|\leq t_2\}
\\&-\frac{1}{2}P\{\min(Y,Z)\leq t_1,|s-W|\leq t_2\}.
\end{align*}
After some calculations we get
\begin{align*}
\xi^{\ex}(s,t_1,t_2)&=
    \frac{1}{2} e^{-s - 2 t_1 -
  t_2} \big(-1 + e^s + e^{t_2} (-e^s + e^{t_2}) I\{s > t_2\}\\& -
   e^{t_1} I\{
     s > t_1\} (e^s - e^{t_1} +
      e^{t_2} (-e^s + e^{t_1 + t_2})I\{
        s > t_1 + t_2\})\big)
\end{align*}
Now we calculate the variances of these projections  $\sigma_K^2(t_1,t_2)$ under $H_{0}.$
\begin{align*}
\sigma_{\ex}^2(t_1,t_2)&=\frac{1}{12} e^{-5 t_1 - 3 t_2}\Big((-1 + e^{t_1})^2 (-1 - e^{t_1} + e^{t_2} + e^{2 t_2})I\{t_1\leq t_2\}\\
&+(-1 + e^{t_2})^2 (-1 +e^{t_1} -e^{t_2} + e^{2 t_1})I\{t_1>t_2\}\Big)
\end{align*}
The plot of this function is shown in Figure \ref{fig: sigmaK1}.

We find that
\begin{equation*}
\sigma_{\ex}^2=\sup_{ t_1,t_2\geq0} \sigma_{\ex}^2(t_1,t_2)=\sigma_{\ex}^2(0.453,0.669)=0.0223.
\end{equation*}
Therefore,  our family of kernels $\Xi^{\ex}(X,Y,Z,W,t_1,t_2)$  is non-degenerate. The arguments about the asymptotic distribution are analogous to the previous cases.

\section{Bahadur efficiency of the proposed tests}
In this section we calculate Bahadur efficiency of the proposed tests with respect to some common alternatives.
First, we give a brief summary of Bahadur's theory. 

The Bahadur efficiency can be expressed as the ratio of
Bahadur exact slope, function describing the rate of exponential decrease for the
attained level under the alternative, and double Kullback-Leibler distance between null and alternative distribution.
More details on Bahadur theory can be found in  \cite{bahadur1}, \cite{nikitinKnjiga}.

 The Bahadur exact slopes are defined as follows.
 Suppose that the sequence  $\{T_n\}$ of test statistics under alternative  converges in probability to some finite function $b(\theta)$.
 Suppose also that the following large deviations limit
 \begin{equation}\label{ldf}
  \lim_{n\to\infty} n^{-1} \ln
P_{H_0} \left( T_n \ge t  \right)  = - f(t)
 \end{equation}
   exists for any $t$ in an open interval $I,$ on which $f$ is
continuous and $\{b(\theta), \: \theta > 0\}\subset I$. Then the Bahadur exact slope is
\begin{equation}\label{slope}
c_T(\theta) = 2f(b(\theta)).
\end{equation}
The exact slopes always satisfy the inequality
\begin{equation}
\label{Ragav}
c_T(\theta) \leq 2 K(\theta),\, \theta > 0,
\end{equation}
where $K(\theta)$ is the Kullback-Leibler distance between the alternative $H_1$ and the class of distributions with densities $\{g_{\lambda}(x)\}$ 
defined by null hypothesis $H_0$, i.e.

\begin{equation}\label{kldef}
K(\theta) = \inf_{\lambda>0} \int_0^{\infty} \ln [g(x,\theta) / g_\lambda(x)] g(x,\theta) \ dx.
\end{equation}

In view of (\ref{Ragav}), the local Bahadur efficiency of the sequence of statistics ${T_n}$ is naturally defined as
\begin{equation}\label{localBahadurEf}
e^B (T) = \lim_{\theta \to 0} \frac{c_T(\theta)}{2K(\theta)}
\end{equation}

\bigskip
 The local Bahadur efficiency is measured for alternative distributions that are close to the null. Therefore we define the following class
 of alternatives.

 Let $\mathcal{G}=\{G(\cdot,\theta)$, $\theta \geq 0\}$, be a family of distributions with densities $g(\cdot,\theta)$,
  such that $G(\cdot,0)$ belongs to the null family of distributions, and the regularity conditions from (\cite{nikitinKnjiga}, Chapter 6) hold.
  Denote  $h(x)=g'_{\theta}(x,0)$. It is obvious that $\int_{-\infty}^{\infty}h(x)dx=0$.

\bigskip Now we return to our test statistics.

In order to apply the Theorem \ref{glavna}  we need to show that the condition of monotonicity in parameter holds for our test statistic.
We shall show it holds for statistic $K^{\ex}_n$, for the others it is analogous and simpler.

Similarly to \cite{nikitinLDKS}, let us divide the intervals $[a_1,b_1]$ and $[a_2,b_2]$ into $N$ parts with the following nodes $t^{k}_{i,N}=\log\frac{N}{N-i}$, $k=1,2$, 
$i=1,...,N$. The nodes are taken such that $N$ parts have probability $1/N$ under null hypothesis.
Put $T_{i,j,N}=[t^{(1)}_{i,N},t^{(1)}_{i+1,N}]\times[t^{(2)}_{j,N},t^{(2)}_{j+1,N}]$, $i,j=1,2,...,N$
For fixed $i,j$
\begin{align*}\sup\limits_{(t_1,t_2)\in T_{i,j,N} }|H_n^{\ex}(t_1,t_2)-G_n^{\ex}(t_1,t_2)|&\leq G_n^{\ex}(t^{(1)}_{i+1,N},t^{(2)}_{j+1,N})-H_n^{\ex}(t^{(1)}_{i+1,N},t^{(2)}_{j+1,N})\\&+H_n^{\ex}(t^{(1)}_{i+1,N},t^{(2)}_{j+1,N})-H_n^{\ex}(t^{(1)}_{i,N},t^{(2)}_{j,N}).
\end{align*}
Thus
 \begin{equation*}
 \Delta_n(N)=H_n^{\ex}(t^{(1)}_{i+1,N},t^{(2)}_{j+1,N})-H_n^{\ex}(t^{(1)}_{i,N},t^{(2)}_{j,N}).
 \end{equation*}
 We have
 \begin{align*}
 \Delta_n(N)&=\frac{1}{\binom{n}{2}}\sum\limits_{1\leq r\leq s}I\{\min(X_r,X_s)\in [t^{(1)}_{i,N},t^{(1)}_{i+1,N}],|X_r-X_s|\in [t^{(2)}_{j,N},t^{(2)}_{j+1,N}]\}\\&\leq
 \frac{1}{n^2}\sum\limits_{1\leq r\leq s}I\{\min(X_r,X_s)\in [t^{(1)}_{i,N},t^{(1)}_{i+1,N}]\}.
 \end{align*}
 Let $\tau_N $ be the sequence of real numbers that converges to zero. Put $\mathcal{T}_i=[t^{(1)}_{i,N},t^{(1)}_{i+1,N}]$. Then
 \begin{align*}
 P\{\Delta_N>\tau_N\}&\leq P\Big\{\frac{1}{\binom{n}{2}}\sum\limits_{1\leq r\leq s}I\{\min(X_r,X_s)\in \mathcal{T}_i\}\geq \tau_N\Big\}\\
 &\leq P\Big\{\frac{1}{\binom{n}{2}}\sum\limits_{1\leq r\leq s}I\{X_r\in\mathcal{T}_i\}\geq \tau_N\Big\}+ P\Big\{\frac{1}{\binom{n}{2}}\sum\limits_{1\leq r\leq s}I\{X_s\in\mathcal{T}_i\}\geq \tau_N\Big\}\\&\leq
  P\Big\{\frac{2}{n-1}\sum\limits_{1\leq r\leq s}I\{X_r\in\mathcal{T}_i\}\geq \tau_N\Big\}\\
  &= P\Big\{\sum\limits_{r=1}^nI\{X_r\in\mathcal{T}_i\}\geq \frac{n-1}{2} \tau_N\Big\}.
 \end{align*}
 Since the summands have the same distribution and are independent the sum has a binomial distribution  $\mathcal{B}(n,\frac{1}{N})$. Applying the inequality from \cite{nikitinLoss} and putting $\tau_N=(\log N)^{-\frac{1}{2}}$ (see also \cite{nikitinLDKS}), for sufficiently large $N$ we get
 \begin{equation*}
 P\{\Delta_N > \tau_N\}\leq e^{-(n-1)\sqrt{\log N}}.
 \end{equation*}
 
 Therefore the condition of the monotonicity in parameter holds.
 
 \bigskip
 
Since our three kernels are non-degenerate, centered and bounded we can find the large deviation function from \eqref{ldf} using Theorem \ref{glavna}.
We present them together in the following lemma.
\begin{lemma}
 \label{ldlema}
 Let $\varepsilon>0$. The large deviations for statistics $K^{\pa}_n$, $K^{\lo}_n$ and $K^{\ex}_n$ are all analytic for sufficiently small $\varepsilon$ and the admit the following
 representations:
 \begin{itemize}
 \item \begin{equation*}
f_{\pa}(\varepsilon)=2\varepsilon^2+o(\varepsilon^2),\; \varepsilon\rightarrow 0,
\end{equation*}
\item \begin{equation*}
f_{\lo}(\varepsilon)=5.87\varepsilon^2+o(\varepsilon^2),\; \varepsilon\rightarrow 0,
\end{equation*}
  \item \begin{equation*}
f_{\ex}(\varepsilon)=0.715\varepsilon^2+o(\varepsilon^2),\; \varepsilon\rightarrow 0.
\end{equation*}
 \end{itemize}
\end{lemma}

In the following lemma we derive the limit in probability of our test statistics under alternative hypotheses.

\begin{lemma}\label{bTeta}
For a given alternative density $g(x;\theta)$ whose distribution belongs to $\mathcal{G}$

\begin{equation}\label{bPar}
 b_{\pa}(\theta) = 2\theta \sup\limits_{t_1,t_2>1}\Big|\integ\xi^{\pa} (x;t_1,t_2)h(x)dx\Big|+o(\theta), \, \theta \to 0.
 \end{equation}

 \begin{equation}\label{bLog}
 b_{\lo}(\theta) = 3\theta \sup\limits_{t_1,t_2\in \mathbb{R}}\Big| \int\limits_{-\infty}^{\infty} \xi^{\lo} (x;t_1,t_2)h(x)dx\Big|+o(\theta), \, \theta \to 0.
 \end{equation}

 \begin{equation}\label{bExp}
 b_{\ex}(\theta) = 4\theta \sup\limits_{t_1,t_2>0}\Big| \int\limits_{0}^{\infty} \xi^{\ex} (x;t_1,t_2)h(x)dx\Big|+o(\theta), \, \theta \to 0.
 \end{equation}

 \end{lemma}
\noindent\textbf{Proof.} We prove only \eqref{bPar}. The others are analogous.

Using Glivenko-Cantelli theorem for $U$-empirical distribution functions \cite{helmersjanssen} we have

 \begin{align*}
 b_{\pa}(\theta)&= \sup_{t_1, t_2 >1}|P\{X>t_1,Y>t_2\}-P\{X>t_1t_2\}| \\&= \sup_{t_1, t_2 >1}\Big|\int\limits_{t_1}^{\infty}\int\limits_{t_2}^{\infty}
 g(x;\theta)g(y;\theta)dxdy-\int\limits_{t_1t_2}^{\infty}g(x;\theta)dx\Big|.
 \end{align*}
 Denote $a_{\pa}(\theta)=\int\limits_{t_1}^{\infty}\int\limits_{t_2}^{\infty}
 g(x;\theta)g(y;\theta)dxdy-\int\limits_{t_1t_2}^{\infty}g(x;\theta)dx$.
 It is easy to see that $a_{\pa}(0)=0$. The first derivative of $a_\pa(\theta)$ along $\theta$ at $\theta=0$ is

 \begin{align}
  \nonumber a_{\pa}'(0)&=\int\limits_{t_1}^{\infty}\int\limits_{t_2}^{\infty}
 h(x)y^{-2}dxdy+\int\limits_{t_1}^{\infty}\int\limits_{t_2}^{\infty}
 h(y)x^{-2}dxdy\int\limits_{t_1t_2}^{\infty}h(x)dx \\\nonumber &=
 2\integ h(x)\Big(\frac{1}{2t_2}I\{x>t_1\}+\frac{1}{2t_1}I\{x>t_2\}-\frac{1}{2}I\{x>t_1t_2\}\Big)dx \\&=
 2\integ h(x)\xi^{\pa}(x)dx.\label{aprim}
 \end{align}

 Expanding the function $a_{\pa}(\theta)$ in Maclaurin's series we obtain \eqref{bPar}.
 \hfill$\Box$

 In what follows we shall calculate the local Bahadur efficiency of our tests for some alternatives.

 \subsection{Statistic $K_n^{\pa}$}

 The alternatives we are giong to use are the following
 \begin{itemize}
  \item a mixture alternative with density
 \begin{equation}
 \label{mixture}
 g(x,\theta)=\frac{1-\theta}{x^2}+\frac{\beta\theta}{x^{\beta+1}}, x> 1,\theta \in (0,1)
 \end{equation}
 \item a Ley-Paindaveine alternative with density
 \begin{equation}
 \label{LayP2}
g(x,\theta)=\frac{1}{x^2} - \frac{\pi \theta \cos\big(\pi (1 - \frac{1}{x})\big)}{x^2},\; x> 1,\theta \in \big(0,\frac{1}{\pi}\big)
 \end{equation}
 \end{itemize}

 The double Kullback-Leibler distance for close alternatives can be expressed as (see \cite{obradovic})

 \begin{equation}\label{klpar}
 2K^{\pa}(\theta)=\theta^2\bigg(\int\limits_{1}^\infty x^{2}h^2(x)dx-
 \bigg(\int\limits_{1}^{\infty} h(x)\ln x dx\bigg)^2\bigg)+o(\theta^2).
\end{equation}

For the mixture alternative \eqref{mixture} we get that the function from \eqref{aprim} is

\begin{align*}
 a'_{\pa}(0)&=\frac{1}{2(t_1t_2)^{1+2\beta}}\Big(I\{t_1<t_2\}(t_1^{1+2\beta}t_2^{\beta}+t1^{2\beta}t_2^{1+\beta}-(t_1t_2)^{1+\beta}-(t_1t_2)^{2\beta}\\&+
 t_1^{1+\beta}t_2^{2\beta}-t_1^{1+2\beta}t_2^{\beta})+
 I\{t_1>t_2\}(t_1^{\beta}t_2^{1+2\beta}+t1^{1+\beta}t_2^{2\beta}-(t_1t_2)^{1+\beta}\\&-(t_1t_2)^{2\beta}+
 t_1^{2\beta}t_2^{1+\beta}-t_1^{\beta}t_2^{1+2\beta})\Big).
\end{align*}

\begin{figure}[h!]
\begin{center}
\includegraphics[scale=0.8]{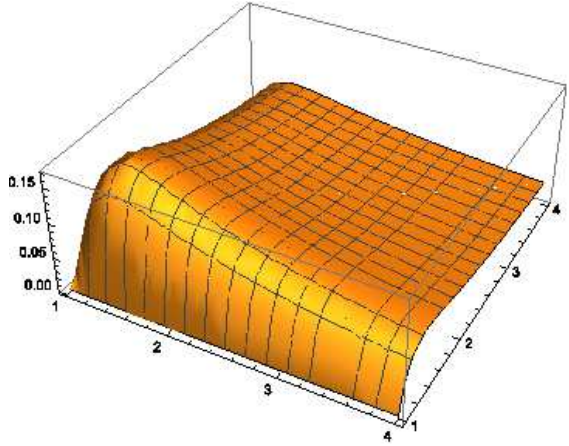}\caption{Plot of the function $a'_{\pa}(0)$, mixture alternative with $\beta=6$}
\label{fig: mixture6}
\end{center}
\end{figure}

 For $\beta=6$ its plot is given in Figure \ref{fig: mixture6}.

 The supremum is attained at the point (1.43,1.43) and it is equal to 0.170.
 The double Kullback-Leibler distance is 1.58. Using \eqref{localBahadurEf}, Lemmas \ref{ldlema} and \ref{bTeta} and \eqref{klpar}
 we obtain that the local Bahadur efficiency is 0.29.

 For the second alternative using same reasoning we get that the local Bahadur efficiencies is 0.23.

 \subsection{Statistic $K_n^{\lo}$}
For logistic distribution there are no standard alternatives so we consider the following
\begin{itemize}
\item a shifted logistic distribution with density
\begin{equation}
\label{logisticlocation}
g(x,\theta)=\frac{e^{-(x-\theta)}}{(1+e^{-(x-\theta)})^2},\;\;x\in R,\;\theta\in (0,1),
\end{equation}
\item a generalized logistic distribution (GLD) with density
\begin{equation}\label{GLD}
g(x,\theta)=\frac{(1+\theta)e^{-x}}{(1+e^{-x})^{2+\theta}},\;\;x\in R,\;\theta\in (0,1).
\end{equation}
\end{itemize}

In case of the family of logistic distribution in the expression \eqref{kldef} it is not possible to find the infimum analytically. Therefore we cannot derive a general expression similar to \eqref{klpar}
and we must calculate it for each alternative separately.  Using the theorem on implicit function to solve extremal problems we find that in the case of alternative \eqref{logisticlocation} we have that $\widetilde{\lambda}(\theta)$ that minimizes the Kullback-Leibler distance is $\widetilde{\lambda}(\theta)=1+o(\theta),\;\;\theta\to 0$,
 while in the case of alternative \eqref{GLD} we have that this minimum is attained for $\widetilde{\lambda}(\theta)=1-0.35\theta+o(\theta)$. Inserting these values into \eqref{kldef} and expanding it into  Maclaurin series we get that double Kullback-Leibler distances for small $\theta$ are $0.33\theta^2+o(\theta^2)$ and $0.82\theta^2+o(\theta^2)$ respectively.

\begin{figure}[h!]
\begin{center}
\includegraphics[scale=0.8]{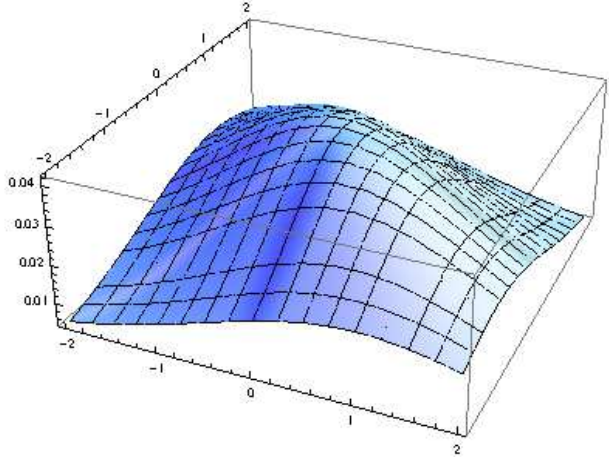}\caption{Plot of the function $a'_{\lo}(0)$, shifted logistic alternative}
\label{fig: shiftLog}
\end{center}
\end{figure}

For shifted logistic alternative the function $a'_{\lo}(0)$ analogous to \eqref{aprim} is too complicated to display. Its plot is given in Figure \ref{fig: shiftLog}.

The supremum is attained at the origin and it is equal to 0.0417.

Using \eqref{localBahadurEf}, Lemmas \ref{ldlema} and \ref{bTeta} and the expression for the double Kullback-Leibler distance from the above paragraph, we obtain that the local Bahadur efficiency for this alternative is 0.55. In case of the generalized logistic alternative the same reasoning produces 0.43.

\subsection{Statistic $K_n^{\ex}$}

As alternatives to exponential distribution we consider two standard alternatives
\begin{itemize}
\item a Makeham alternative with density
\begin{equation}
 g(x,\theta)=\big(1+\theta(1-e^{-x})\big)e^{-x-\theta( e^{-x}-1+x)},\theta > 0, x\geq 0
\end{equation}
\item a Weibull alternative with density
\begin{equation}
 g(x,\theta)=(1+\theta)x^\theta e^{-x^{1+\theta}},\theta > 0, x\geq 0.
\end{equation}
\end{itemize}

It can be shown (\cite{NikTchir}) that for small $\theta$ equation \eqref{kldef} can be expressed as
\begin{equation}\label{kul}
2K^{\ex}(\theta)=\bigg(\int\limits_{0}^{\infty }h^2(x)e^xdx-\Big(\int\limits_{0}^{\infty}xh(x)dx\Big)^2\bigg)\cdot \theta^2+o(\theta^2).
\end{equation}

In case of Makeham alternative we get  

\begin{align*}
 a'_{\ex}(0)&= \frac{1}{6}e^{-3t_1-2t_2}(-1+e^{t_1})(-1+e^{t_2}).
\end{align*}

\begin{figure}[h!]
\begin{center}
\includegraphics[scale=0.8]{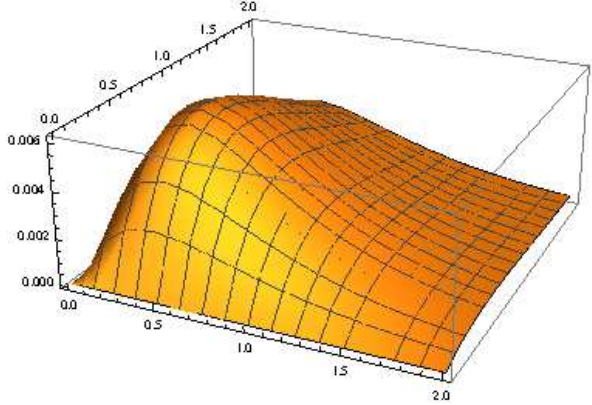}\caption{Plot of the function $a'_{\ex}(0)$, Makeham alternative}
\label{fig: makeham}
\end{center}
\end{figure}

Its plot is given in Figure \ref{fig: makeham}.

The supremum is attained at the point (0.405,0.693) and it is equal to 0.00617.

Using Using \eqref{localBahadurEf}, Lemmas \ref{ldlema} and \ref{bTeta} and \eqref{kul} we find that the local Bahadur efficiency is 0.38. In case of Weibull alternative the same procedure gives the efficiency of 0.20.

\section{Conclusion}

In this paper we proposed and analyzed three tests for three different distributions based on different types of characterizations.
They all have two-dimensional Kolmogorov-type statistics and are consistent against any alternative. They are also free of corresponding parameter $\lambda$ which enables us to test the composite null hypotheses. We calculated their local Bahadur efficiencies against some
alternatives. To be able to do this we gave a general large deviation theorem that could be applied to our statistics. The efficiencies are reasonable and comparable to
some other Kolmogorov-type tests based on characterizations.

\end{document}